\documentclass[prd,apd,twocolumn,nofootinbib]{revtex4}

\usepackage{graphicx}
\def\gr{$\gamma$-ray}

\begin{document}
\title{PeV neutrinos from interactions of cosmic rays with the interstellar medium in the Galaxy}

    \author{A. Neronov$^{1}$,
          D.Semikoz$^{2}$,
          C.Tchernin$^{1}$
          }
  \affiliation{$^1$ISDC, Department of Astronomy, University of Geneva, Ch. d'Ecogia 16, 1290, Versoix, Switzerland 
            \\$^2$APC, 10 rue Alice Domon et Leonie Duquet, F-75205 Paris Cedex 13, France \\
}

\begin{abstract}
We present a self-consistent interpretation of the of very-high-energy neutrino signal from the direction of the inner Galaxy, which is a part of the astronomical neutrino signal reported by IceCube. We demonstrate that an estimate of the neutrino flux in the $E>100$~TeV energy range lies at the high-energy power-law extrapolation of the spectrum of diffuse \gr\ emission from the Galactic Ridge, as observed by Fermi telescope. This suggests that IceCube neutrino and Fermi/LAT \gr\ fluxes are both produced in interactions of cosmic rays with the interstellar medium in the Norma arm and/or in the Galactic Bar. Cosmic rays responsible for the \gr\ and neutrino flux are characterised by hard spectrum with the slope harder than $-2.4$ and cut-off energy higher than $10$~PeV.  
\end{abstract}
\maketitle

\section{Introduction}

\begin{figure*}
\includegraphics[width=0.7\linewidth]{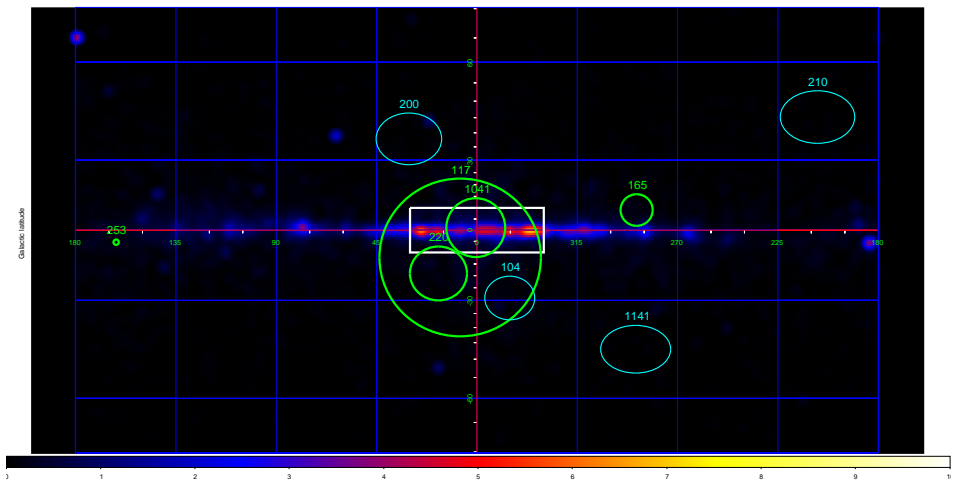}
\caption{Fermi count map in the energy range above 100 GeV, smoothed with $4^\circ$ Gaussian.   Elliptical regions show the arrival direction and its uncertainty of the IceCube neutrino events with energies above 100~TeV. Green colour marks neutrinos coming from directions close to the Galactic Plane. Numbers next to each ellipse mark the energy of neutrino events. White rectangle shows the spectral extraction region for the Galactic Ridge. }
\label{fig:skymap}
\end{figure*}

IceCube collaboration has recently reported detection of two PeV energy neutrino events which constitute an excess of  $2.8\sigma$ over background \cite{Aartsen:2013bka} .  In addition, a search  of neutrino signal in the energy band above 30 TeV was performed  and an evidence of the astrophysical signal  on top of the background of atmospheric neutrinos and muons was reported at the level of $4.1\sigma$ \cite{icecubeIPA}. The highest energy events are mostly cascade events in the IceCube detector. Because of this, measurement of the directions of the incoming neutrinos suffers from poor angular resolution, in the 10 degrees range. Nevertheless, some of the 28 high-energy events belong to event clusters, with the strongest cluster of 5 events detected in the direction of the central part of the Galactic Plane.  

The central part of the Milky Way galactic disk is the strongest \gr\ source on the sky \cite{fermi_diffuse}. Most of the \gr\ emission from the Galaxy is produced by the cosmic ray interactions with the interstellar medium, which result in production and subsequent two-photon decays of neutral pions. The same interactions also lead to  neutrino emission from the Galaxy, via production and decays of charged pions \cite{Stecker79,Berezinsky:1992wr}. The neutrino and \gr\ signals from the cosmic ray interactions are expected to have comparable flux and energy distributions. Therefore, the Galaxy is expected to be (one of)  the strongest astronomical source of high-energy neutrinos. 

The Galaxy taken as a whole is a very extended astronomical source on the sky. This poses a problem for the search of neutrino signal from this source. Indeed, signals from isolated point sources could be distinguished from the background of atmospheric neutrinos and muons as "excess" events coming from some particular sky directions. To the contrary,  it might be difficult to separate the very extended neutrino signal from the Galaxy from still more extended atmospheric neutrino and muon signal, or from an isotropic extragalactic neutrino flux based on the spatial morphology of neutrino arrival directions. Still, the central part of the Milky Way and, in particular the central part of the Galactic Plane is generically expected to be a brighter neutrino source than e.g. the outer Galaxy. This distinguishing feature of the Galaxy source could be used to identify the Milky Way contribution to the astronomical neutrino flux, in particular of the flux detected by IceCube at the energies above 30~TeV.

\begin{figure}
\includegraphics[width=\linewidth]{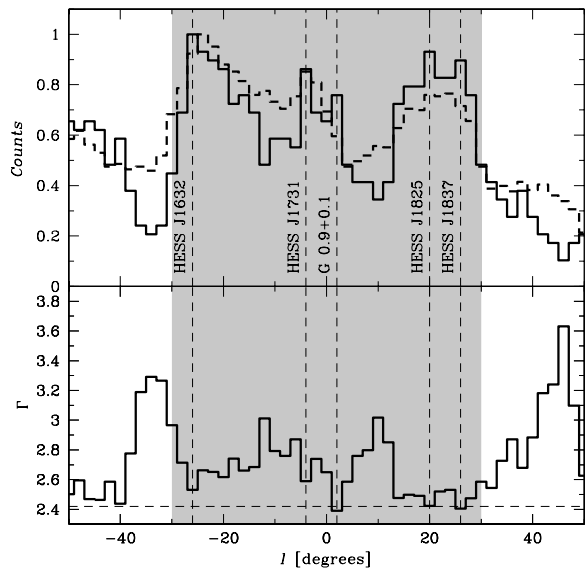}
\caption{Top: Normalised intensity profile of \gr\ emission from $8^\circ\times 8^\circ$ regions centred at a given $l$ position along the Galactic Plane. Solid line shows the profile in the 0.3-1~TeV band. Dashed line is for the 0.1-0.3~TeV band. Bottom panel shows the slope of the spectrum calculated from the ratio of the counts in the two bands. Shaded range shows the boundaries of the Galactic Ridge spectral extraction region. Vertical dashed lines mark positions of brightest sources.} 
\label{fig:profile}
\end{figure}

Detection of multi-TeV neutrino signal from the Galactic Plane would be important because this would trace the locations of recent injections of cosmic rays in the interstellar medium \cite{candia05,Neronov:2012kz}. These recent events of injections of cosmic rays might be traced by a range of extended (several degree scale) \gr\ sources dominating the emission from the Galaxy in the very-high-energy band  \cite{Neronov:2012kz}. The observed statistics of hard extended \gr\ sources is consistent with that expected in the model in which  the observed cosmic ray content of the Galaxy is produced by episodic injections by supernova explosions over the last 10-30~kyr. This provides an (almost) direct evidence for the hypothesis of association of cosmic ray injection with supernova explosion events. However, firm identification of the hard extended sources around recent supernova explosions as the places of  injections of cosmic rays could be achieved only via detection of neutrino flux from these sources. 

The conjecture that Galactic sources of TeV-PeV cosmic rays should reveal themselves at the first place through the extended (rather than point source) \gr\ and neutrino emission is motivated by a recent progress in the study of spreading of cosmic rays in the interstellar medium. Cosmic ray diffusion on 100~pc scale distance range is found to proceed in a very anisotropic manner, with much faster diffusion along certain directions, as compared to isotropic diffusion, even in case of pure  turbulent magnetic field~\cite{Giacinti:2012ar,Giacinti:2013wbp}. The same is true for the diffusion even at kiloparsec scale distances in the presence of a regular component of magnetic field (e.g. in the Galactic arms). Those findings suggest that neutrinos and \gr\ signal produced by TeV-PeV cosmic rays injected by young (tens of kiloyears old) sources should primarily come from large multi-degrees angular scale regions.

In this paper we follow a previously developed approach of Ref. \cite{tchernin13} and combine the analysis of gamma-ray data of Fermi Large Area Telescope (LAT)  in inner Galaxy and new IceCube data\footnote{Note that a the analysis used for the 28 high-energy events \cite{icecubeIPA}, sensitive to the shower events produced by electron, tau neutrinos  and neutral current interaction has much larger effective area in the Southern hemisphere than the standard IceCube analysis based on muon tracks. This explains the improvement of sensitivity of IceCube to the signal from the inner Galactic Plane, compared to the estimate of Ref. \cite{tchernin13}}. We find that the flux contained in IceCube neutrino events near the Galactic Centre lies at the high-energy extrapolation of the \gr\ spectrum of the inner part of the Galactic Disk (from -30 to 30 degrees Galactic longitude range).  Based on this observation, we argue that cosmic ray interactions in the Galactic Ridge should be considered as a viable source of the observed 0.1-1~PeV neutrinos from the inner Galaxy. 

The suggested measurement of the neutrino flux from  the Galactic Ridge, if confirmed with deeper IceCube exposure, would have strong implications for the origin of cosmic rays with hard spectrum, responsible for the production of the hard component of the \gr\ emission. Detection of PeV neutrino from this region imposes a lower bound on the cut-off energy $E_{cut}\gtrsim 10$~PeV of the hard cosmic ray spectrum. This is consistent with an observation of  ankle-like structure in the spectrum of the "light" cosmic rays by KASCADE-Grande~\cite{KASCADE-Grande} together with recent limits on cosmic ray anisotropy by Auger Observatory~\cite{Auger_anisotropy}, which  imply that transition from Galactic to  extra-galactic cosmic rays starts only above $10^{17} $ eV \cite{galactic_to_extragalactic}. This, in turn,  means that at least some Galactic sources  should accelerate cosmic ray protons well beyond the knee.

\section{\gr\ and neutrino data analysis.}

The inner part of the Galactic Plane is the brightest \gr\ source on the sky. The \gr\ emission originates from cosmic ray interactions in the interstellar medium. The same interactions produce also neutrinos and it is natural to expect that the inner part of the Galaxy is also one of the brightest neutrino sources on the sky. 

The neutrino signal at the energies below 100~TeV is dominated by the atmospheric muon and neutrino background. The IceCube analysis reported in the Ref. \cite{icecubeIPA} shows that some 70\% of detected events in the energy band 30-100~TeV are due to the atmospheric background. In the energy band above 100~TeV the background event fraction is below 10\%. Taking this into account, we concentrate in the following on  the events with energies $E>100$~TeV.

Fig. \ref{fig:skymap} shows the distribution of arrival directions of the highest energy ($E>100$~TeV) neutrinos detected by IceCube, superimposed onto a Fermi \gr\ image of the sky in the energy band above 100~GeV. One could see that among 9 neutrino events, three are distributed around the Galactic Ridge, a bright \gr\ emission region in the Galactic longitude range $-30^\circ<l<30^\circ$. Two more events are arriving from the directions close to the Galactic Plane and four events are arriving from high Galactic latitude. 

Low statistics of events and absence of clustering of events at high Galactic latitudes and in the outer Galactic disk lead to a too large range of possible viable models for the origin of those neutrinos. The high Galactic latitude events could be produced by e.g. extragalactic sources, like blazars \cite{tchernin13a}, while the events from the Galactic Plane may be from individual Galactic sources or be a part of diffuse emission from the Galaxy. 

In principle, the same applies also to the neutrinos coming from the inner Galaxy. However, taking into account the coincidence of arrival directions of a subset of $E>100$~TeV neutrinos with the direction toward the inner Milky Way disk, we put forward a conjecture that the Galactic Ridge is a source of those neutrinos. To verify the self-consistency of this conjecture, we estimate the $E>100$~TeV neutrino flux and compare it with an extrapolation of the measured \gr\ spectrum of the Galactic Ridge. If both \gr s and neutrinos are produced via the same mechanism (cosmic ray interactions resulting in pion production and decays), the neutrino flux and spectrum is expected to be nearly identical with the the \gr\ flux and spectrum. 

In an alternative model in which  neutrinos are produced by individual isolated sources in the inner Galaxy or beyond it, there should be no particular reason for the consistency of the spectral characteristics of the \gr\ diffuse emission from the Galactic Plane with those of the neutrino signal. Instead, if the neutrino signal originates from a particular isolated source, the \gr\ spectrum of this source should be consistent with the flux / spectrum of neutrinos coming from the direction of the inner Galaxy. 

The sensitivity of \gr\ telescopes in 100~TeV energy band is currently not sufficient for detection of astronomical sources and neutrino flux in this band could be estimated via extrapolation of the \gr\ spectrum from lower energies. The highest 100~TeV band flux is expected from the brightest region with the hardest \gr\ spectrum. Fig. \ref{fig:profile} shows a scan of the spectral properties of the \gr\ emission from the $8^\circ\times 8^\circ$ regions at different Galactic longitudes. The grey band in this Figure marks the Galactic Ridge region $-30^\circ<l<30^\circ$  (see Fig. \ref{fig:skymap}). 

\begin{figure}
\includegraphics[width=\linewidth]{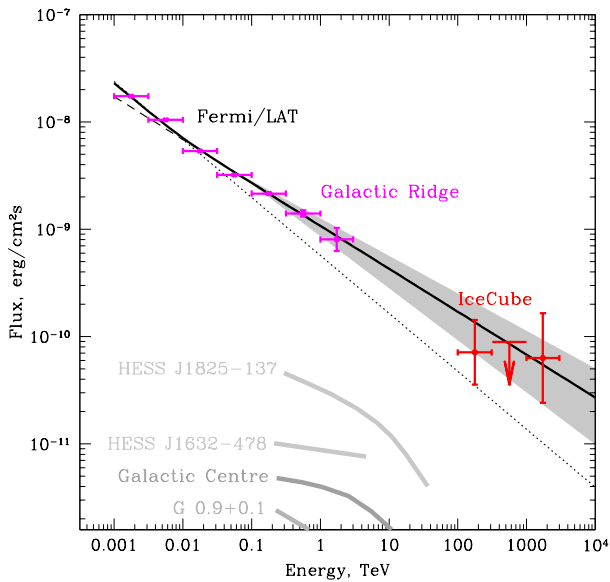}
\caption{Comparison of Fermi/LAT and IceCube spectra of sources in the direction of the inner Galaxy. Magenta data points show the overall \gr\ spectrum of a $-30^\circ<l<30^\circ$, $-4^\circ<b<4^\circ$ part of the Galactic Plane. Red data points show the estimates of IceCube neutrino flux above 100~TeV. Black thick solid line shows a broken power law model for the \gr\ spectrum with soft (thin,dotted) and hard (thin dashed) components. Grey band shows the uncertainty of the spectrum of the hard component.  }  
\label{fig:spectrum_scan}
\end{figure}

From Fig. \ref{fig:profile} one could see that the highest flux  in the range of Galactic longitudes $-30^\circ <l<30^\circ$ is measured from the parts of the Galactic Plane centered at the positions of the sources HESS J1837-069 (an unidentified extended source), HESS J1825-137 (a pulsar wind nebula), a source close to the Galactic Centre supermassive black hole and/or G 0.9+0.1 composite supernova remnant and an unidentified extended source HESS J1632-478. To identify the most probable candidate(s) for the sources of observed neutrinos, we have combined the measured spectrum of \gr\ emission from the hard / bright sources consistent with the IceCube event cluster and and estimate of the neutrino flux from the IceCube data. The result is shown in Fig. \ref{fig:spectrum_scan}. 

The TeV band spectra of individual sources contributing to the Galactic Ridge emission are shown by grey curves.
The overall spectrum of emission from the Galactic Ridge is well represented by a broken power law with a hardening in the 20~GeV range. The harder component of the spectrum is characterised by the photon index $\Gamma\simeq 2.4$, while the softer component has the spectrum with the slope $\Gamma\simeq 2.5$.

To calculate the estimate of the flux and spectrum suggested by the IceCube data we used the following procedure. 
We consider at the first place only the events with energies above 100~TeV. This is done based on the modelling presented in the IceCube publication \cite{icecubeIPA}, from which it follows that the all-sky spectrum below at least 60~TeV is consistent with the expectations from the atmospheric muon and neutrino backgrounds. At the same time, comparison of the background with the observed signal in the energy range above 100~TeV shows that the observed flux is dominated by the astronomical source signal, with only minor / negligible background contribution. 

There are three neutrinos with energies above 100~TeV in the Galactic Centre region event cluster (see Fig. \ref{fig:skymap}). The effective area of IceCube still grows rapidly in the energy range above 100 TeV and the difference between the effective area at 200 TeV and 1 PeV is by a factor of three (i.e. half-an-order of magnitude). Taking this into account, we divide the energy range 100~TeV -- 3 PeV onto three energy bins of equal width in the logarithmic scale: 100-316~TeV, 316~TeV-1~PeV and 1-3.16~PeV. In these energy bins we calculate the effective area from the data presented in \cite{icecubeIPA}. The data given in \cite{icecubeIPA} provides a sky-averaged effective area for electron, tau and muon neutrinos. To find the effective area corresponding to the range of declinations of interest we use the declination dependence of the count rate from an isotropic source \cite{icecubeIPA} and find that the effective area of the detector is almost declination-independent in the Southern hemisphere for the event selection chosen for the analysis. This effective area is by a factor $1.3$ higher than the $4\pi$-averaged affective area reported in \cite{icecubeIPA}. This information is sufficient for the estimate of the flux in the 100~TeV-3~PeV energy band shown in Fig. \ref{fig:spectrum_scan}.

\begin{figure}
\includegraphics[width=\linewidth]{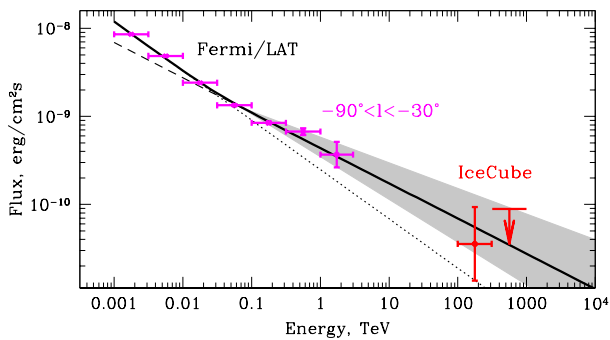}
\caption{Same as in Fig. \ref{fig:spectrum_scan}, but for the region $-90^\circ<l<-30^\circ$. }  
\label{fig:spectrum_l300}
\end{figure}

From Fig. \ref{fig:spectrum_scan} one could see that the IceCube flux estimate lies right at the power law extrapolation of the \gr\ spectrum of the Galactic Ridge to the 100~TeV energy range. At the same time, the estimate of the neutrino flux is inconsistent with the extrapolations of the spectra of individual sources contributing to the Galactic Ridge.   This suggests a model in which the hard component of the \gr\ flux from the entire Galactic Ridge and the neutrino flux from the inner Galaxy direction are produced via one and the same mechanism: interactions of cosmic rays with the interstellar medium.  

The relation between the \gr\ and neutrino signal from cosmic ray interactions in the interstellar medium should hold not only in the Galactic Ridge region, but everywhere along the Galactic Plane. This means that the extrapolation of the \gr\ signal from the sub-TeV toward higher energies should provide a good estimate for the 100~TeV neutrino signal all along the Galactic Plane \cite{tchernin13}. To verify the self-consistency of the model in which the observed $E>100$~TeV neutrinos at low Galactic latitudes are coming from the cosmic ray interactions, we also extract the \gr\ spectrum and estimate the neutrino flux from the region $-90^\circ<l<-30^\circ$, see Fig. \ref{fig:spectrum_l300}. This region is entirely contained in the Southern hemisphere, so that our estimate of the IceCube exposure is also applicable for this region. From Fig. \ref{fig:spectrum_l300} one could see that the detection of one $E>100$~TeV neutrino from the direction toward this part of the Galactic Plane is consistent with the expectations based on the extrapolation of the \gr\ spectrum. 

Counting the numbers of photons with energies above 100~GeV coming from the inner and outer Galaxy, we find that the \gr\ flux from the outer part of the Galactic Disk ($90^\circ<l<270^\circ$) is approximately three times lower than the flux from the inner Galactic Disk \cite{fermi_diffuse}. If both \gr s and neutrinos coming from the direction of the Galactic Plane are produced by cosmic ray interactions, the ratio of neutrino flux from the outer Galactic Plane to that from the inner Galactic plane is also expected to be approximately $1\div 3$. This is what is observed in the IceCube data (see Fig. \ref{fig:skymap}). There are four neutrinos with energies above 100~TeV from the inner Galactic Disk and one from the outer Galactic Disk. This demonstrates the self-consistency of the hypothesis that low Galactic latitude astronomical neutrinos with energies above 100~TeV detected by IceCube could be a part of diffuse emission from the Galaxy. 

Modelling of diffuse \gr\ emission from the inner Galaxy \cite{fermi_diffuse} suggests that a significant part of $E>100$~GeV flux could be due to inverse Compton emission from electrons. In this case, the flux of pion decay contribution to the \gr\ flux measured by Fermi is lower than the total flux of the Galactic Ridge  region shown in Fig. \ref{fig:spectrum_scan}. The spectrum of the pion decay component is softer in the $E\sim 100$~GeV band, so that the estimate of the neutrino flux in the $E>100$~TeV band  is inconsistent with the high-energy extrapolation of the pion decay component of the \gr\ spectrum. Our analysis suggests that an alternative model for the $E>100$~GeV \gr\ emission from the Galaxy might be valid, with the pion decay component dominating up to the highest energies and inverse Compton emission giving only a sub-dominant contribution to the spectrum. 


\section{Discussion}

In this paper we have considered a possibility that  the $E>100$~TeV neutrino signal observed from the direction of the  Galactic Center region is due to the cosmic ray interactions in the inner Galaxy. We have compared different possibilities for the Galactic neutrino source(s): diffuse emission from the entire Galactic Ridge or emission from isolated individual sources. 

From Fig. \ref{fig:spectrum_scan} is is clear that 
the  neutrino flux is not produced by any individual source in the Galactic Centre region, like the Galactic Center itself, G 0.9+0.1 HESS J1632-478 and HESS J1825-137. One could see that the neutrino flux in the 100~TeV energy band exceeds the \gr\ fluxes from individual sources by more than an order-of-magnitude. This mismatch between the \gr\ and neutrino fluxes could not be explained by possible absorption of the highest energy \gr s in the source because (a) normalisation of the flux at lower energies is already much lower than the neutrino flux estimate and (b) all the sources, as well as the interstellar medium around the sources are expected to be transparent for the 100~TeV \gr s: the density of the soft photon backgrounds is largely insufficient for significant attenuation of the \gr\ flux due to the pair production \cite{moskalenko}. 

This suggests that the observed neutrino flux, similarly to the Galactic diffuse \gr\ emission, originates from an extended region in the interstellar medium. Extended emission from the entire Galactic Disk scale height is natural to expect in a scenario in which most of the cosmic rays escape from a source to the interstellar medium and diffuse preferentially along the ordered magnetic field in the Galactic arms. We have demonstrated the self-consistency of such scenario by showing that the  estimate of the neutrino flux agrees with the extrapolation of the \gr\ spectrum of the entire Galactic Ridge spanning the longitude range $-30^\circ<l<30^\circ$. It is remarkable that the estimated neutrino flux in the IceCube event cluster is close to both the 35-year-old \cite{Stecker79} and recent \cite{candia05} estimates of the expected neutrino flux from the Galactic Ridge region.

 The Galactic Ridge spans the range of Galactic longitudes delimited by the projections of the innermost Norma arm of the Galaxy and also of the Galactic Bar  \cite{churchwell09}.  Cosmic rays spreading from sources  in the Norma arm and /or in the Galactic Bar could not spread into $|l|>30^\circ$ range, they only extend along the arm. This might explain the distinct appearance of the brighter Galactic Ridge in the \gr\ map and should lead to a similar extension of the neutrino source in the inner Galaxy (a conjecture which could be verified with a deeper IceCube exposure).

More generally, our analysis is consistent with a general expectation that  strongest PeV  neutrino signal from Galaxy should come from all the Galactic Disk, rather than from isolated point sources.  
Cosmic rays with energies  $E>1$ PeV propagate in intermediate regime between diffusion and linear escape, producing  complicated spatial distributions around their sources, with higher density in the Galactic arms, see e.g. \cite{galactic_to_extragalactic}. 
 Recent detailed calculations of 1-100 PeV cosmic ray propagation in Galaxy \cite{cosmic_ray_knee} have shown that cosmic rays typically escape from Galactic disk in 200 kyr
at PeV energies and in about 20 kyr at 10 PeV energies and above. At 10 kyr time scale 10 PeV cosmic ray protons spread on several kpc scales in direction of regular fields and just on 0.5 kpc in perpendicular directions. Thus all proton sources in Galactic Ridge would contribute to observed  neutrino spectrum. 
1-10 PeV cosmic rays injected by by the sources in the Norma arm and in the Galactic Bar  spread over the angular distance range $\theta\simeq R/D\simeq 20^\circ-30^\circ$ on 20~kyr time scale. Thus, unless there is a large population of sources much younger than 10~kyr, the cosmic ray induced neutrino signal is expected to come from the entire Galactic arms, rather than from isolated sources.


Of course, the statistics of the neutrino signal is at the moment too low to make firm conclusions on the identification of the Galactic Ridge as the source of the observed neutrinos. Alternative possibilities are not ruled out. It is possible that the signal in the direction of the inner Galaxy is produced by the cosmic ray interactions in the more extended Fermi bubble structures \cite{ahlers,razzaque}, or a distribution of cosmic ray sources like hyper nova remnants \cite{liu}, or by the  extragalactic sources \cite{winter,anchordoqui}, see \cite{anchordoqui}  for a review. Further accumulation of statistics of neutrino events from the inner Galaxy will confirm or ruled out the model proposed in this paper in a straightforward way.

{\it Acknowledgement}. We would like to thank F.Halzen for useful discussion and comments on the text. The work of AN and CT is supported by the Swiss National Science Foundation. The work of DS is supported in part by the grant RFBR 13-02-12175-ofi-m.

\end{document}